# Scintillation Event Localization in Hemi-Ellipsoid Detector for SPECT, a simulation study using Geant4 Monte-Carlo

Hanif Rauf Soysal[2], Joyoni Dey[1a], William Patrick Donahue[3], Kenneth L. Matthews[1]

[1]Department of Physics and Astronomy, LSU, Baton Rouge, LA, 70803, USA.

[2]University of Mississippi Medical School, Department of Radiation Oncology, Jackson, MI, 39213

[3]Yale University, Department of Therapeutic Radiology, New Haven, Connecticut, 06511

Corresponding author: [1a]J. Dey, PhD, Contact e-mail: deyj@lsu.edu.

**ABSTRACT**

**Purpose:** A high sensitivity Cardiac SPECT system using curved crystals with pinhole collimation was proposed previously (Dey, *IEEE TNS* 2012). Using hemi-ellipsoid CsI crystals, in simulations, the system improved sensitivity by a factor of 3 over newer generation SPECT systems (DSPECT or GE Discovery) while keeping the resolution comparable (Bhusal et al, *Med Phys*, 2019). In this work, we hypothesize that the high curvature detector results in measurable differences in light distribution from events at different depths in the crystal. We rigorously test this by analyzing the scintillation light using Monte-Carlo (Geant4) and propose a statistical event localization method in a hemi-ellipsoid detector. We evaluate the localization error of the algorithm and the back-projected errors in object space. **Methods:** To develop this localization capability for the proposed design, we used Geant4 to simulate the propagation of scintillation light in a monolithic hemi-ellipsoidal CsI crystal. A look-up table (LUT) was created to map the points inside the crystal to the expected light pattern on the crystal surface using the Geant4 simulation data. Thirteen zones were considered across the crystal. In each zone, gamma-rays were simulated and the resulting photon intensity on the surface was captured, serving as our experimental interactions. An algorithm based on Poisson statistics was developed to limit the search of the experimental gamma-ray event locations into smaller regions of the LUT. The localization was fine-tuned by comparing the light distribution of the gamma interactions in selected patterns from the LUT points and then recorded. The algorithm-localized gamma-ray events were also individually back-projected to the object mid-plane, (expected mid-plane of the heart), and the error at the plane was recorded as well. **Results**: The light patterns of adjacent LUT points showed visually discernible differences. Excluding some outliers (up to 2%), the localized errors averaged over all the zones was **0.71 (+/-0.44) mm** with a worse case of 1.36 (+/-0.67) mm at the apex. Moreover, when back-projected to the midplane of the region of interest for Cardiac SPECT, the errors were < *1mm* due to the high system magnification afforded by the apex and other zones. The average back-projection error at the mid-plane of the object was **0.4mm +/-0.22mm. Conclusion:** We modeled gamma-event interactions and scintillation light spread in CsI hemi-ellipsoid detector and developed a robust statistical algorithm that localized scintillation events to within 0.71(+/-0.44) mm on the average within a hemi-ellipsoid CsI detector. Moreover, due to the high system magnification afforded by the crystal, the hemi-ellipsoid unit was capable of achieving **<1mm** average localization in the object space, assuming perfect pinhole collimator resolution recovery.

Thus, we show that this high sensitivity system will be able to deliver images with high resolution for Cardiac SPECT. In the future, the application of this may be extended to Brain SPECT and small animal imaging.

**INTRODUCTION**

Single Photon Emission Tomography (SPECT) is a functional imaging modality that is used for cardiac imaging applications to assess myocardial perfusion, ischemic defects, and abnormal heart wall motion. About ~7 million patients/year undergo nuclear cardiology scans in the USA. However, of all the diagnostic imaging modalities, nuclear medicine is the second highest contributor of radiation exposure to the general public, behind Computed Tomography (CT)[1-3]. Dey et al. have investigated SPECT systems with curved detectors[4-6] which provides increased sensitivity (thus operable with a lower input dose) without loss of resolution. For example[6], 21 inverted wine-glass sized hemi-ellipsoid detectors with pinhole collimators can increase the sensitivity ~3.35 times over the new generation of dedicated Cardiac SPECT systems such as GE Discovery and DSPECT. The reconstructed resolution after resolution recovery is comparable or better than these systems, at an average of 4.44mm in the region of interest. This allows for ultra-low-dose imaging (3 mCi) at ~5 minutes for comparable clinical counts as the state-of-the-art system.

While other investigators have used curved detectors before for PET[7] or SPECT[8], they are typically larger crystals contouring the body. The proposed design (US patent[5], Dey[4], Bhusal et al[6]) is different in that the detectors are smaller (inverted wine-glass) with high curvature. They can afford high magnification for pinhole-collimation. In prior work[6], we conservatively assumed the detector resolution to be 3mm.

Depth-of-interaction localization that exists in PET/SPECT[10-19], in conjunction with readout innovations and advancements, has greatly improved the accuracy of these modalities. Using Monte-Carlo modeling and depth-of-interaction estimation based on statistical matching, Maas et al[10] achieved 1.05mm FWMH in the detector composed of a 20cm (flat) crystal with APD readout on the frontal surface of crystal (surface exposed to the gamma-ray). Schaart et al[11] achieved ~1.56-1.58mm FWHM resolution at the center of a PET detector using SiPM readout for frontal surface and about ~4-4.23mm for back-surface readout of a 10mm (flat) crystal. For SPECT, back-surface-readout is commonly used due to higher attenuation of gamma-rays by frontal surface readout and the use of these methods is sparse in clinical SPECT, particularly for monolithic crystals. However, in a recent DOI correction for monolithic crystals for SPECT by Bettiol et al.[14] an empirical model for light-distribution (by J.W. Scrimger et al. (1967)[16] ) was used for a flat-crystal slab achieving top-vs-bottom discrimination (3mm) for a 6mm thick $LaBr_3$:Ce.

In this work, we hypothesize that the high curvature of the hemi-ellipsoid crystal will create unique light patterns depending on depth-of-interaction, and a sophisticated detector model similar to that used in PET-literature[13,10-11] will allow for accurate depth-of-interaction correction. Therefore, we investigate the scintillation light spread using Geant4 Monte-Carlo[9] within the hemi-ellipsoid detector of 6mm for SPECT application. In what follows we performed comprehensive modeling of the light spread in Monte-Carlo (Geant4) to create a look-up-table (LUT) and then developed a statistical algorithm to narrow down our event-localization search and then fine-tuned the localization.

**METHODS**

*1. Brief Rationale*

A detector with high curvature relative to its thickness may help localize events within the crystal through the variation in light distribution with depth and position of interaction. For example, the scintillation events, A and B, are at two depths approximately along the same normal of the crystal in **Figure 1**. Notice how an event at point A has a limited line of sight to the apex of the crystal, which reduces the amount of light. However, an event at point B can be expected to deposit more light on the photosensors near the apex assuming the refractive indices of the crystal, light-guides, and APD are well matched so that internal reflection is limited. As we show in our simulation results, we can utilize the light spread to our advantage to distinguish depth-of-interaction.

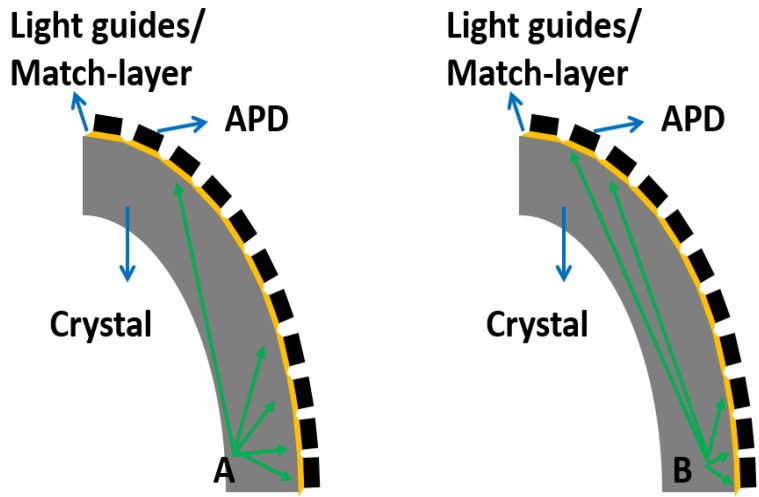

**Figure 1**. Schematic diagram of two scintillation events at different depths (not to scale)

## 2. Geant4 Simulation for Constructing Look-up-Table

We adopt a Look-Up-Table (LUT) approach to determine the point of interaction for the photons in the crystal. The LUT is created by precomputing light profiles for all the points in the crystals (ideally). Assuming interactions at each unique location in the crystal result in a unique intensity distribution on the detector's surface within statistical noise, it should be possible to determine the point of interaction, S, for a measured intensity distribution by finding its closest match in the LUT. The closer the resemblance, the better we can pinpoint the true location of the unknown interaction. This approach is illustrated in **Figure 2** for a hypothetical LUT on a flat crystal as an illustration of concept.

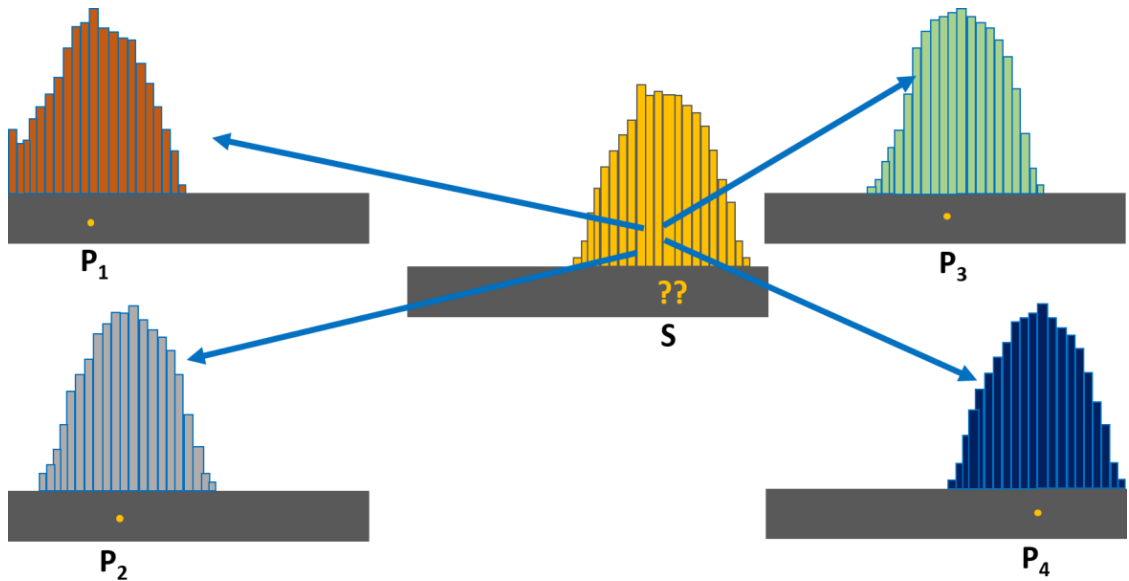

**Figure 2:** Hypothetical Depiction of the LUT approach. The four histograms P1-P4 showing photon counts in binned positions are precomputed forming the LUT in a section of the crystal (shown in grey). When an unknown light pattern S is obtained for image acquisition, the binned light pattern (shown in yellow) can be compared to the light patterns in the LUT to localize S.

In this example scenario, points $P_1$ through $P_4$ represent LUT distributions at different locations in a flat scintillating crystal (shown in gray, repeatedly for convenience). An event at each of these locations might produce the scintillation light patterns shown as photon count histograms above the crystal. During image acquisition or experiment, when a gamma-ray interacts with the detector, light pattern S (middle), is obtained via the photosensors. S can be compared to the light patterns in the LUT, with the best match(es) allowing us to localize the experimental light pattern. In this case, it can be determined that the experimental event S must have occurred near $P_4$ because the light pattern is the most similar. All the surrounding patterns can then be used to fine-tune the localization. A robust non-linear algorithm is described later for these steps.

This approach can be extended to our 3D hemi-ellipsoid crystal, allowing us to use leverage the variations in light distribution created by interactions caused by the non-constant rate of curvature for the crystal. By accurately modeling what the light patterns look like in 3D, we can then determine the approximate location of future scintillation events in 3D.

*2.1 Geant4 Model*

We used Geant4 Monte Carlo[9] to simulate the illumination of the outer surface of the CsI crystal by scintillation events that occurred at various chosen points in the crystal. The following Physics processes associated with photons were enabled: optical absorption, optical Rayleigh scattering, scintillation, Cerenkov, and diffuse and specular reflection and refraction boundary processes.

To be able to control the location where scintillation events occurred for the formation of the LUT, we simulated photoelectric absorption events directly at the chosen points, instead of directing gamma-rays that interact at random locations along the path inside the crystal. Only absorption events are considered to create the LUT since this is the most likely interaction. However, when evaluating LUT localization for gamma-rays, both scatter events (as well as absorption events) will be considered later (in Section 3 within Methods).

We considered that the absorption of a 140.5 keV gamma-ray would produce on average 9132 photons[20]. This number was arrived at by considering CsI(Tl) produces about 65000 photons on the average per MeV (Knoll[20], chart Pg. 238). Thus a 140.5keV gamma-ray produces $65000 \times 0.1405 \approx 9132$ photons on average. For each point used to create the LUT, 100 such scintillation events were simulated, for a total of 913,200 photons emitted isotopically at each point. This number was deemed statistically accurate as results were identical to the third decimal place when 1,000 events were used as a test.

The crystal was constructed as a hemi-ellipsoid shell (further referred to as simply "ellipsoid"). The Z-axis of the crystal goes through the geometric center of the full ellipsoid and the very tip of the apex. The X- and Y- axes form the plane which encompasses the base of the crystal. The outer surface of the crystal has a base in the XY-plane with an outer radius of 46 mm and an inner radius of 40 mm (so that the crystal is 6 mm thick at the base). The outer surface of the crystal extends to a height of 126 mm in the Z-direction toward the apex and the inner surface extends to 120 mm (again, so that it is 6 mm thick at the apex). A secondary "photosensor" or ellipsoid shell was created on top of the existing crystal ellipsoid for the actual detecting of the incoming photons. The passing of the photon into this new detecting medium was considered a detection event and the simulated scintillation photon was stopped upon entry into this detecting medium, representing 100% collecting efficiency. A small ring was created at the base of the crystal to act as a reflective coating for the base so that photons hitting the bottom edge of the crystal were reflected. The geometry produced in Geant4 is shown in Error! Reference source not found.**3**. Though not shown in the simulations, the detector design has a pinhole collimator attached at the base. This hypothetical pinhole extends 60 mm below the base of the crystal.

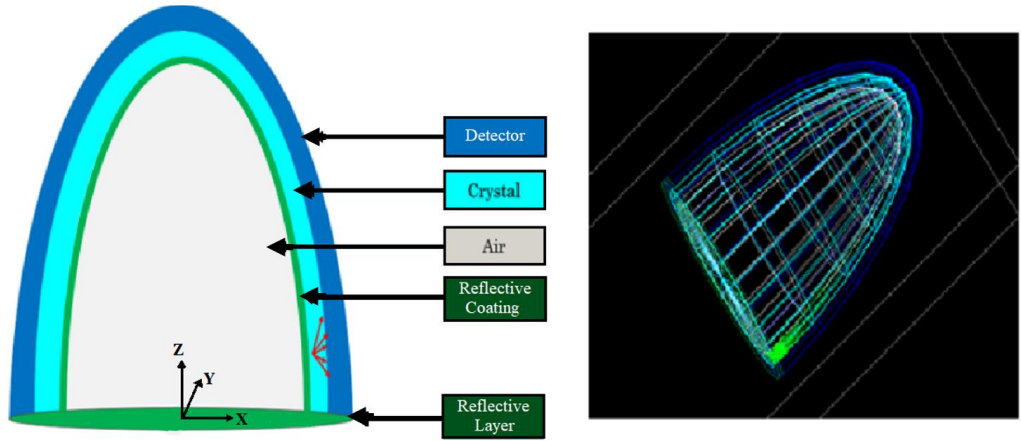

**Figure 3.** Hemi-Ellipsoid Detector Geometry in Geant. The Dark blue ellipsoid represents the photosensors, i.e., the detecting element, the cyan represents the crystal, and the white represents the inside of the ellipsoid. A green ring at the bottom is present for reflection from the bottom (annular) edge of the crystal. At the lower right part, the paths of scintillation photons in the crystal being simulated are shown (in red). The co-ordinate axes are shown in the bottom, with z-x as central vertical cross-section and y-axis into the plane of the paper.

For the simulation, the UNIFIED model[13] was used with a dielectric-dielectric surface (as this represents the CsI-to-epoxy surfaces of the photosensor elements) and a dielectric-metal surface (as this represents the CsI-to-reflective-coating on the inner surface and base of the crystal). The surface between the crystal and the outer photosensors was set with a "ground" finish. An Epoxy layer was used for impedance matching, and the base and inner surface were simulated as rough ground and coated in Teflon to mimic the reflective coating. Note that Teflon has a high reflectance (>95%) for the wavelengths under consideration. This reflectance is automatically accounted for in the Geant4 simulations. So, even though we started with 9132 photons, after reflectance the number of photons will reduce from 9132, varying with wavelength. The crystal was modeled by setting its material to containing 99.6% CsI and 0.4% Tl[21]. A refractive index database was used to obtain the refractive indices at the desired energies, which cites the Journal of Physical and Chemical Reference Data[22-23]. Data from CERN's Crystal Clear Collaboration was used to obtain the absorption lengths of the crystal at various energies[24]. A standard plot[25] of the gamma-ray energy-frequency distribution produced by the scintillation of a 140.5keV gamma-ray in CsI(Tl) was used. Reflectivity for the crystal was set to 100% diffuse reflection.

*2.2 Look-Up-Table Points*

To precisely obtain low localization error, the LUT sampling was chosen as 1mm. Due to the rotational symmetry of the crystal, only the interaction points for a single slice of the crystal need to be simulated to represent all possible light distributions. To achieve this without introducing sampling/interpolation issues, we selected the slice of the crystal on the XZ plane at Y=0. To further reduce the computational burden, only points with X≥0 were simulated for the LUT points, again invoking rotational symmetry.

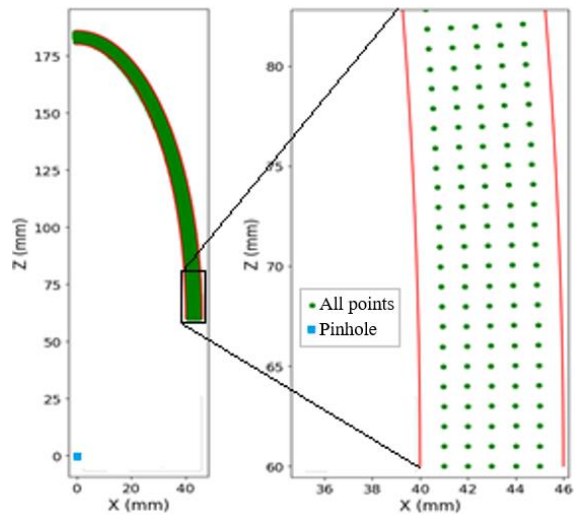

**Figure 4.** All LUT points that can be generated with at most 1 mm spacing in the central slice of the crystal in the XZ plane. Left: All points in the main slice. Right: Zoomed-in image. The location of a hypothetical pinhole is shown at the origin of the image on left.

The location of each LUT point was found using normal vectors calculated at 1mm intervals (Euclidean distance) along the outer surface of the crystal. Then, along each normal vector from the outside edge to the inside edge of the crystal, LUT points were calculated in 1mm intervals as well. Effectively, this ensured that each point in the crystal is *at most 1mm* away from the neighboring points (since the 1-mm-apart inward normal vectors on the outer edge are converging). The full set of points that will be simulated along one slice is shown in **Figure 4**. There were 715 such points. Only a subset of these points was simulated, for considerations of computation, based on the zones that would be tested, as discussed in the next section. To generate the complete 3D LUT for the crystal, each LUT point (the subset of the 715 that was chosen) was rotated by 1mm about the central axis. For each LUT point that was rotated, the scintillation photons that were simulated to have hit the outer surface of the crystal were also rotated by the same angle. This is justified by the rotational symmetry arguments discussed earlier. While detailed later, it is worth mentioning that the light measured at the outer surface is sampled with 2mm x 2mm bins (example footprint of an APD).

*2.3 Look-Up-Table Binning*

A method was required to bin the data in a way to match as closely-as-possible the bins to the geometry of real-life optical photosensors (which will be added to the outer surface of the crystal during manufacturing). For the purposes of this study, we assumed 2mm by 2mm photosensors, such as APDs, would be placed around the outer surface of the crystal. The photon detector efficiency of the APDs is considered deterministic at 85%. This value is obtained from Mosset[26], for a wavelength around 500 nm. We corrected for a fill factor of 64% (typical) to obtain 0.85 x 0.64 or about 54% overall efficiency.

We generated the bins to discretize the data by creating 2 mm thick (latitudinal) rings from the base to the apex of the crystal. Each of these rings was then split into segments that were roughly 2 mm apart in arc length, with slight adjustments in arc length to make sure the bins go completely around the ring. Each of these bins was then assigned a "key" used to refer to it. When counts would be accumulated in the crystal, the value assigned to this key would be incremented by one. This is illustrated in **Figure 5**. The 2mm x 2mm tiling is approximate at the top-most layer but is adjusted with light guides. Smaller 1 mm x 1 mm APDs may also be used to tile the apex area. We tested this but, in our case, this did not change the results significantly. Hence, we maintained the same-sized APDs throughout the crystal.

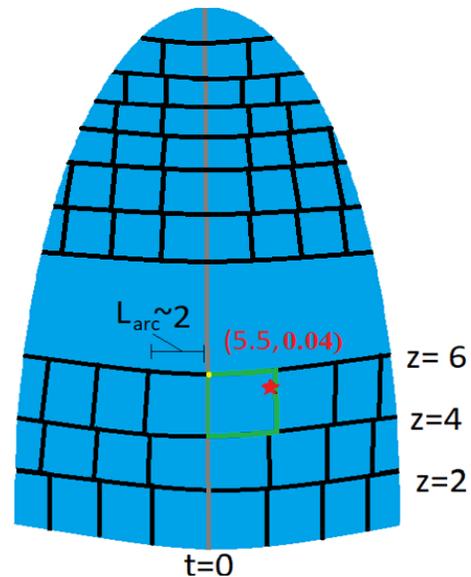

**Figure 5.** 2D Illustration of 3D bins. Each bin has a *key* <Z,θ>, where Z represents the height of the ring and the angle θ is determined by the x- and y-coordinates of the bin. The green-colored bin is represented by the key at the yellow point. An optical photon detected at the red star would result in an increment of the bin designated by that yellow key.

*3. Geant4 Evaluation using Gamma-rays*.

While the LUT was generated by direct scintillation (absorption) events to better control sampling, we used mono-energetic gamma-rays for evaluation. This ensures we consider scatter as well as absorption events. The following physics processes were considered for the gamma-ray: Compton Scattering and Photoelectric Effect. Gamma-rays were directed at the crystal in 13 angular intervals (zones) from the location of a hypothetical pinhole (in the physical design of the system, 60 mm below the base of the crystal) to assess the performance of the crystal as shown in **Figure 6**. The zones are numbered 1 through 13 from apex to base. Not all LUT points were simulated due to considerations of computation. Only LUT points within a large neighborhood of these gamma-ray paths were simulated. At the apex, the zones are spaced finer (translating to about a 5° angle between zones 1 and 2 as measured from the pinhole and only a handful of LUT points are in between gamma-ray paths 1 and 2). As we move toward the base, the spacing becomes less fine (with up to 14 LUT points in between gamma-ray paths). The spacing is finer near the apex to be able to better examine the different effects caused by the high radius of curvature.

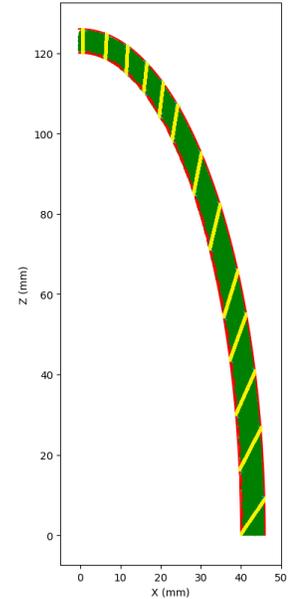

**Figure 6**. Thirteen Zones: Gamma-ray paths sent through crystal
The path of gamma-rays sent through the crystal are shown in cyan.

To determine the path the gamma-rays would take in each of the thirteen zones, a point halfway into the crystal (i.e., 3 mm in depth) and between the rotated LUT and original LUT was chosen. The gamma-rays were generated and sent toward this point, originating from the center of the hypothetical pinhole (the point at (0,0,-60)). 1000 gamma-rays were sent to each of the regions for evaluation. The same binning method was used for these gamma-ray interactions as for the LUT points.

*4. Region-narrowing, and Localization algorithms*

In order to localize interaction events of gamma-rays, we need to compare the light patterns generated by the interaction events with the light patterns of events in our look-up table. However, it is inefficient and time-consuming to compare the interaction events' light patterns with all of the LUT's light patterns. Therefore, a statistical algorithm was first developed to limit the region of the LUT to be used for localizing, i.e. the region-narrowing algorithm.

This algorithm's starting point is to consider the "mode-bin" **MB**—the bin with the largest number of counts—for any LUT point or experimental event. One method to limit the number of LUT points that need searching would be to simply compare the mode-bin of the experimental event **MB$_{exp}$** with the mode-bin of all the LUT points **MB$_{LUT, i}$**. If the mode-bins match, that LUT point can be set aside to be used for the subsequent localization algorithm. However, this method used on its own lacks robustness, due to statistical variation in the signal produced by events in the same location. Therefore, by invoking Poisson statistics, we compared the mode-bin of an experimental gamma-ray **MB$_{exp}$** with a larger set of bins **{B}$_i$**, for each LUT point *i* and not just the mode-bins **MB$_{LUT,i}$** of each LUT point. For each LUT point, this set of bins **{B}$_i$** is chosen so that if an experimental event (of unknown locale) occurred exactly at that LUT point, we can be 95% confident that **{B}$_i$** will contain the experimental mode-bin **MB$_{exp}$**.

The Poisson statistics comes into play when generating **{B}$_i$**. Recall that each LUT point's light pattern is the expected light pattern obtained from averaging 100 runs. By utilizing the average number of counts in the mode-bin (of the 100 runs), as well as the standard deviation of the mode-bin, we can use Poisson statistics to calculate which bins would have an upper bound that would be within two standard deviations of the average in the mode-bin. Thus, a small set of bins **{B}$_i$** becomes associated with each LUT point P$_i$ instead of having just one bin (the **MB**) associated with each LUT point P$_i$.

The method is illustrated in **Figure 7** for a hypothetical 2D crystal. We assume in this example that the average counts in the mode-bin **MB$_{avg}$** for a particular LUT point is 400 and that the standard deviation **MB$_{STD}$** follows Poisson statistics, making it 20. Thus, a 95% confidence lower bound for the modebin would be 360 counts, which becomes the upper bound **MB$_{UB}$** for the bins in **{B}$_i$**. We can then use Poisson statistics and solve for the number of counts, **c**, that a bin must have to be included in **{B}$_i$**. Thus, in this case, all bins with average counts above 324 are included in **{B}$_i$**, as demonstrated in the figure. (To solve for c, we utilized the fact that $MB_{UB} = MB_{AVG} - 2MB_{STD}$ and that $MB_{UB} = c + 2\sqrt{c}$. This allows us to solve for c, giving us: $c = \left(1 - \sqrt{1 + MB_{AVG} - 2MB_{STD}}\right)^2$ or in this case: $c = \left(1 - \sqrt{1 + (400) - 2(20)}\right)^2 = 324$.).

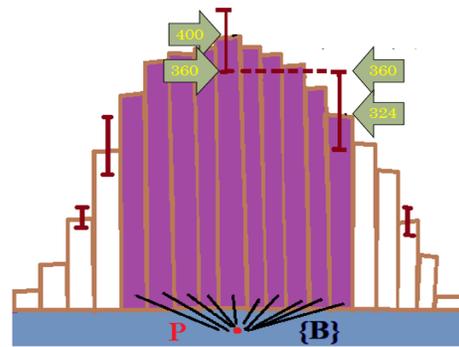

**Figure 7**. Example of identifying region

**Figure 8** further illustrates the method schematically in 2D. All bins from the experimental distribution (S) with unknown locations have been removed except the mode-bin (yellow). Refer back to **Figure 2** for the original schematic displaying the full light distribution. All bins from the other LUT distributions have been removed as well, except for {**B**}$_i$ sets for each LUT point. The number of bins in {**B**}$_i$ for each LUT point is shown exaggerated in **Figure 8** for display clarity. In reality, only a small fraction of all the bins are in {**B**}$_i$ for each LUT point. The mode-bin of the experimental distribution (yellow bar) will be compared against the {**B**}$_i$ of the other four distributions. Notice that for P$_1$ and P$_2$, their {**B**}$_i$ do not overlap the mode-bin, and so they will not be considered in the search algorithm. However, since the {**B**}$_i$ for P$_3$ and P$_4$ overlap the mode-bin of the experimental distribution (at their ends), they will be used in the search algorithm. The same approach is extended to our 3D crystal.

In the small percentage of cases where this algorithm is expected to fail to produce a match, a brute-force method is used. A geographical narrowing-down of the region of LUT points to be searched is found by considering all the LUT points within a 1 cm radius of the point 3 mm deep in the crystal directly under the **MB**$_{exp}$.

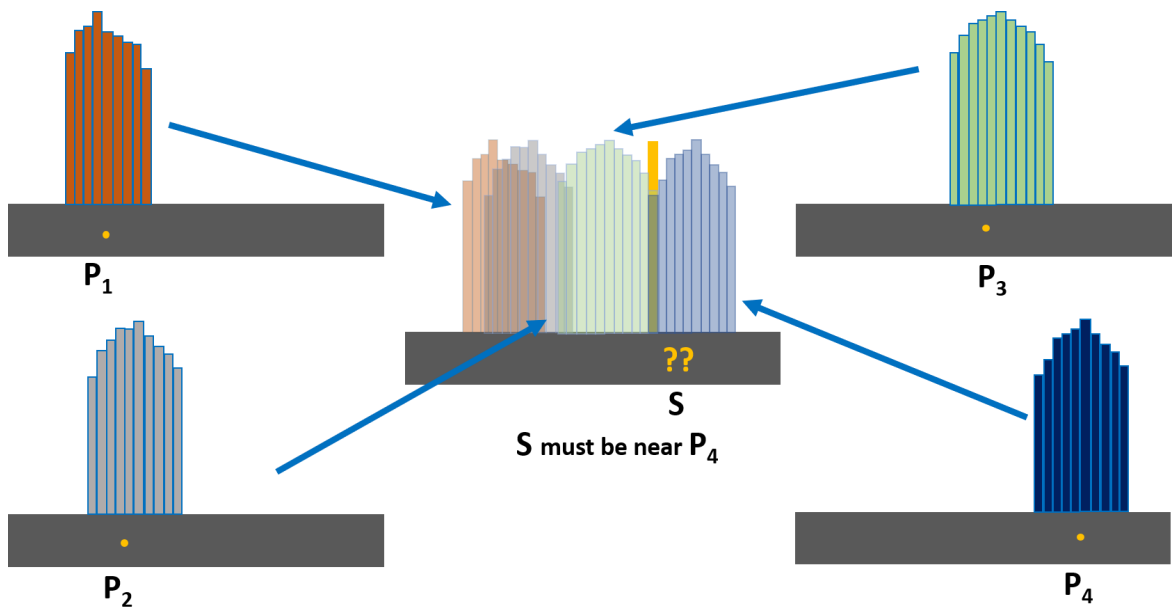

**Figure 8**. Comparison of the **MB**$_{exp}$ (yellow bar) for the experimental point of unknown location in a flat crystal against the {**B**}$_i$ of LUT points. Notice how the {**B**}$_i$ of P$_4$ and P$_3$ overlap with the **MB**$_{exp}$ of S.

To recapitulate, we have reduced our computational time and efficiency while maintaining robustness in the following way: instead of comparing the entire light distribution of an experimental gamma-ray with the light distribution of *every* LUT point, we proposed to compare only the **MB**$_{exp}$ with the **MB**$_{LUT}$ for each point in the LUT. However, we pointed out that comparing only one bin would lack robustness, so instead, we chose to compare **MB**$_{exp}$ with the {**B**}$_i$ (not just the mode-bin), where {**B**}$_i$ is a group of bins for each LUT such that we can be 95% confident that a random interaction event occurring exactly at that LUT location will contain **MB$_{exp}$**. Finally, we put in a fail-safe in the few cases where this algorithm may fail (due to statistics).

After we narrowed down the region of LUT points to search, we employed a bin-to-bin sum-squared error (SSE) to compare the light pattern of the gamma-ray interaction to the light patterns of each of the LUT points in the narrowed-down region. Then, in 3D, interpolation was performed using *the inverse* of at most the eight smallest SSEs as weights for averaging the coordinates of the LUT points. Thus, the smaller the SSE, the more its coordinates were weighted when localizing the gamma-ray interaction. Eight points were chosen because the eight lowest SSEs expected are those from the closest LUT points (i.e., the eight LUT points that would form a cuboid around the point of interaction of the experimental gamma-ray).

## RESULTS

1. *Illustration of Light Spread within Crystal*

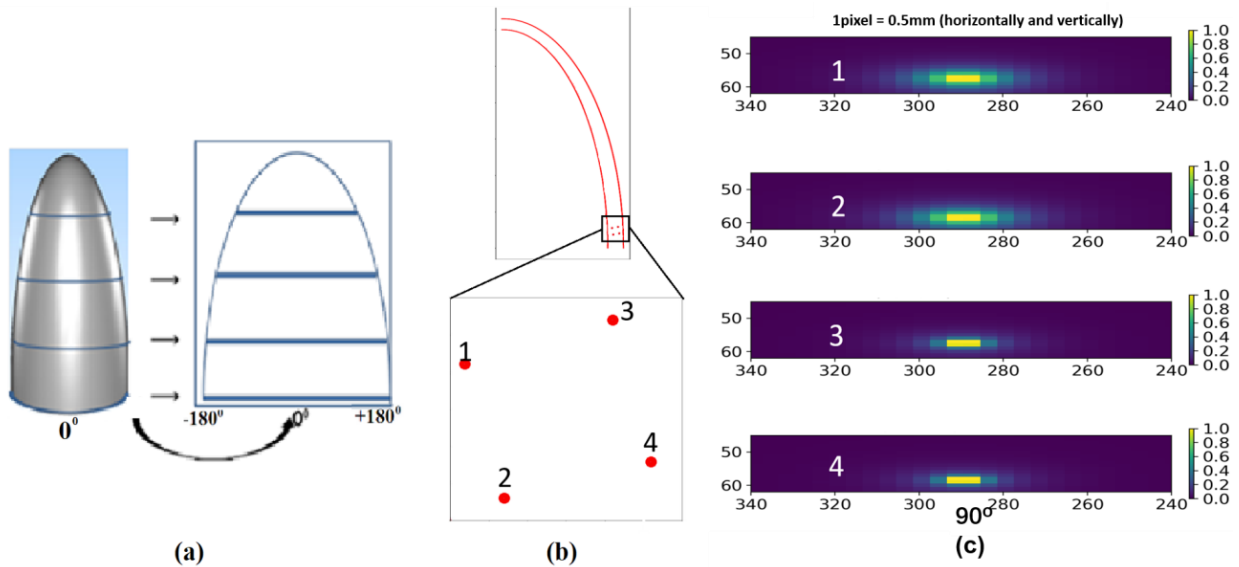

**Figure 9.** The crystal is cut and opened up for 2-dimensional display, as seen in (a) schematically. It is cut at ±180° (which corresponds to ± 144.5mm around the outer base of the crystal). (b) shows the location of four adjacent LUT points near the crystal base. (c) shows their light distributions (with relative intensities) in 2D (zoomed in, as most bins will not have any light). The pixel size in (c) **is 0.5mm,** horizontally and vertically.

To illustrate the 3D light intensity in the bins, we cut open the 3D hemi-ellipsoid and lay it out as shown in **Figure 9a**. **Figure 9b** shows a subset of 4 neighboring LUT points near the base of the crystal, and **Figure 9c** shows a zoomed version of their light distribution. Notice how there are visually discernible differences, especially in-depth, i.e., with points 1 and 2 versus 3 and 4.

## 2. Evaluation with Gamma-rays in Geant4

Depending on the zone, 88%-98% of gamma-rays actually interacted in the crystal, with the percentage of points interacting in the crystal increasing as we moved from the apex (zone 1) to base (zone 13). Of the 88% to 98% of gamma-rays (880 to 980) that interacted inside the crystal, about 110 to 125 (11% to 12.5% of total simulated) of these gamma-rays interacted with a Compton Scatter event. Of the Compton scattered events, about a sixth of these or 15 to 20 (1.5% to 2% of total simulated) scattered very far from the path of the gamma-ray (more than 3 mm, and generally tens of millimeters). Gamma-rays that scattered egregiously far away were excluded from our search algorithms, treated as statistical outliers. The 1.5-2% of the data that are outliers scattered far and are likely to unrealistically skew the errors estimates.

For each gamma-ray in each zone, the difference between the localized point and the location of the actual interaction was recorded. This was done for gamma-rays that didn't scatter (shown in **Figure 10**) as well as for gamma-rays that interacted and may have scattered up to a maximum of 2 mm away from the original gamma-ray path (shown in **Figure 11**). The localized points (used to calculate **Figure 11**) were subsequently back-projected to a plane of z = -210 mm (150 mm below the pinhole as considered in prior work[6], where the object mid-plane is expected to be). The gamma-ray paths were also back-projected to this plane. The difference (or error) between the back-projected algorithm-localized point and the back-projected gamma-ray is shown in **Figure 12**. The figures show the average localization error for the two cases (not including and including Compton-scattered gamma-rays) and those that fall within one and two standard deviations.

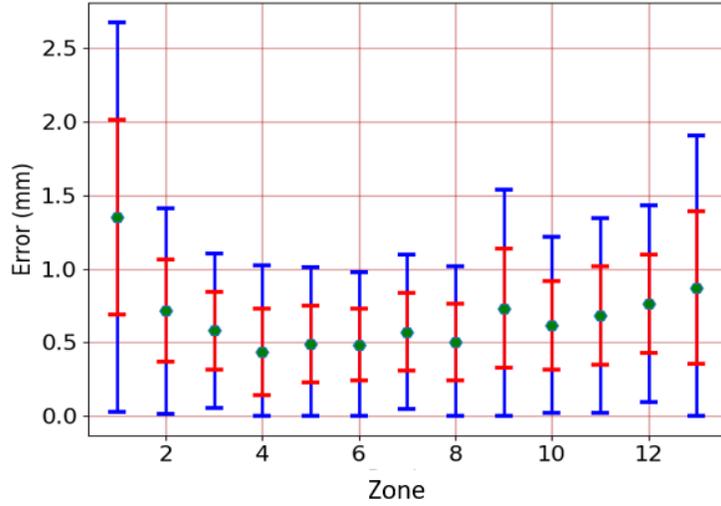

**Figure10**. Localization Error for gamma-rays that didn't scatter but were absorbed within the crystal. Red and blue bars indicate one and two standard deviations, respectively

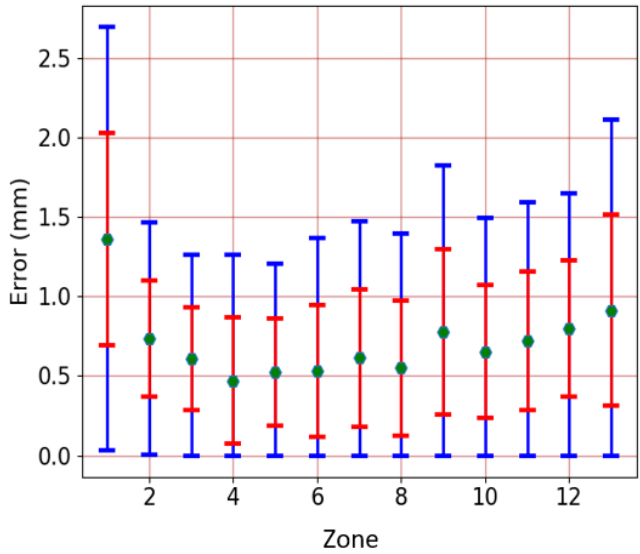

**Figure11**. Localization Error for gamma-rays, including those that scattered and then were absorbed, within 2mm of gamma-ray paths (excluding 2-3% outliers**).** Red and blue bars indicate one and two standard deviations, respectively.

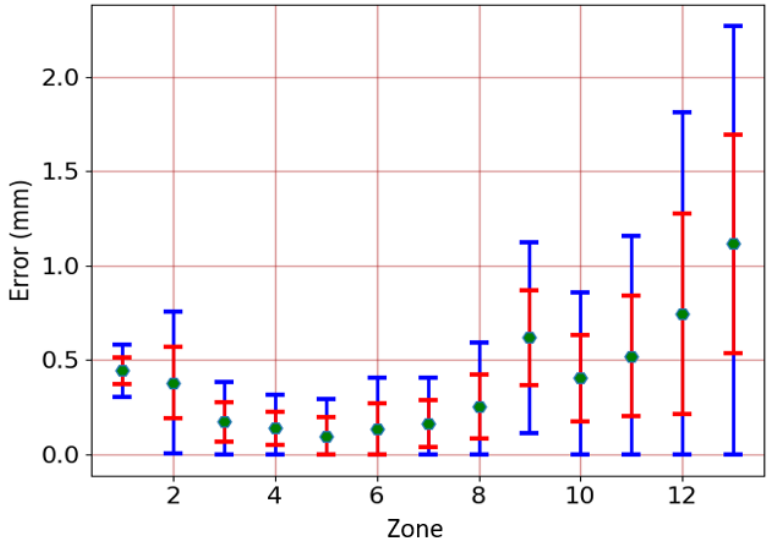

**Figure12**. Error when localized points are back-projected to central plane of object, for points considered in Figure 12. Red and blue bars indicate one and two standard deviations, respectively

The distribution of localization errors without/with the Compton-scattered gamma-rays, as well as the back-projection errors, are shown in **Error! Reference source not found.**. In all zones, 99% and 98% of all errors were within 2 mm when excluding or including Compton scatter events, respectively. One exception was at the apex where only 79% of the errors were less than 2 mm. However, even at the apex, 99.9% of the errors were within 3 mm. Errors were consistently larger in the apex, dropping significantly in the subsequent zones, and increasing slightly toward the base. When including gamma-rays which Compton-scattered, the localization errors were slightly higher with a little more variance. The data for all zones are also summarized in **Table 1.** The average error in the detector (including Compton scatter) is **0.71mm (+/-0.44)** with a worst-case 1.36mm (+/-0.67) at the apex and the average back-projection error from all 13 zones is **0.4mm (+/-0.22).**

**Table 1.** Statistics for localization errors in all zones. Values are in millimeters.

|      | **Error Without Compton (mm)** | | **Error With Compton (mm)** | | **Error Back-projected w/Compton (mm)** | |
|------|------|-------|------|-------|------|-------|
| **Zone** | **Avg** | **Stdev** | **Avg** | **Stdev** | **Avg** | **Stdev** |
| 1 | 1.35 | 0.66 | 1.36 | 0.67 | 0.44 | 0.07 |
| 2 | 0.71 | 0.35 | 0.74 | 0.37 | 0.38 | 0.19 |
| 3 | 0.58 | 0.26 | 0.61 | 0.32 | 0.17 | 0.11 |
| 4 | 0.43 | 0.29 | 0.47 | 0.39 | 0.14 | 0.09 |
| 5 | 0.49 | 0.26 | 0.52 | 0.34 | 0.10 | 0.10 |

| 6  | 0.48 | 0.25 | 0.53 | 0.41 | 0.13 | 0.14 |
| 7  | 0.57 | 0.26 | 0.61 | 0.43 | 0.16 | 0.12 |
| 8  | 0.50 | 0.26 | 0.55 | 0.42 | 0.26 | 0.17 |
| 9  | 0.73 | 0.40 | 0.78 | 0.52 | 0.62 | 0.25 |
| 10 | 0.62 | 0.30 | 0.65 | 0.42 | 0.40 | 0.23 |
| 11 | 0.68 | 0.33 | 0.72 | 0.44 | 0.52 | 0.32 |
| 12 | 0.76 | 0.33 | 0.80 | 0.43 | 0.74 | 0.53 |
| 13 | 0.87 | 0.52 | 0.91 | 0.60 | 1.11 | 0.58 |

**DISCUSSION**

*Results:* The hemi-ellipsoid curved detector geometry helped to achieve <1.4 mm average event localization error at the detector compared to top-vs-bottom discrimination achieved in Bettiol et al[14] for a 6 mm flat detector crystal. Note that our LUT resolution was 1mm and the localization in-detector was less than 1.36 mm+/-0.67, for the worst-case (at the apex). However, the apex also had the highest magnification and near-normality of the incident gamma-ray. This reduced the localization error significantly when back-projected to object-space. The overall localization error from all 13 zones of the detector to 0.4 mm+/-0.22. Thus, the hemi-ellipsoid curved detector we considered provides a combination of better intrinsic resolution as well as higher overall magnification that helps in reducing the average localization error to submm levels.

*Simulations:* While standard values from the literature were selected for multiple parameters, some values, such as the APD fill-factors, could be different in practice changing the light collected by the APDs. However, our algorithm is normalized with respect to counts and which makes it robust to changes in multiplicative scale factors unless the factors are very small. The high brilliance of the CsI(Tl) provides adequate light to make it robust against Poisson changes in counts reaching the APDs. SNR of Poisson statistics also works in favor, since it varies with $\sqrt{(number\ of\ counts)}$. For example, a 15% reduction in light would produce only an 8% drop in SNR.

*LUT:* The LUT was modeled with direct scintillator light to have better control of point sampling. The actual evaluations were done with gamma-rays to access errors due to (1) algorithm localization performance and (2) errors due to scattering in crystal. **Figure 11** shows the true-algorithm error by considering the gamma-rays that did not scatter (but were absorbed by the crystal). **Figure 12** shows the overall error in the crystal, including scattered-and-then-absorbed gamma-rays within the crystal. We eliminated the 2-3% outliers to allow us to better estimate our errors of the remaining 97-98% of points. Some of these photons, about 1.5-2%, were widely scattered and indeed may be back-projected well outside the ROI.

*Statistical Algorithm:* The statistical algorithm proved robust in localizing the experimental gamma-ray (unknown event location) to a cubic region first which we then fine-tuned by applying a normalized sum-squared error to use as an inverse weight to interpolate the location of the gamma-ray. The statistical algorithm failed to find a region match for only a small fraction of cases, with average over zones of 4.6% (worst case 7.5%, Zone 5) where the 1cm geometric search was used.

*Practical Note:* Several vendors in USA and China were able to provide reasonable quotes of curved crystals of the dimensions, providing the expectation that they can be manufactured economically. Some of the vendors will include acrylic light-guide coverage on the outer surface, optically coupled with index-matching optical grease/adhesive. The inner surface may be coated with matted white paint or Teflon (to approximate a Lambertian diffusely reflecting surface for subsequent modeling). The APDs may be coupled with index-matching Meltmount (Cargille Laboratories, NJ, USA) used by Beekman's group[10] or equivalent optical-grade coupling adhesive.

*Importance of modeling of detector intrinsic resolution loss:* If the collimator resolution loss is not recovered in iterative reconstruction the collimator and detector resolution are in quadrature. For a flat-detector system, resolution at a depth "b" under the pinhole with height "a" and diameter "d" is given by $\text{Rsys} = \sqrt{\{\frac{(b+a)d}{a}\}^2 + \{\frac{b}{a}R_I\}^2}$. For the curved detector, it is more complex since the magnification a/b is varying with angle of incident gamma-rays. The system resolution is derived in Dey[4] for system with the half-paraboloid crystals and can be extended straightforwardly to the hemispherical crystal system. More importantly, in the case when the collimator resolution is recovered within the reconstruction algorithm, the detector intrinsic resolution presents the fundamental limit to the resolution of the system. Recovering detector resolution involves careful modeling and is usually not carried out within the reconstruction algorithm.

Our goal here was to investigate localization errors in the detector. We assumed full collimator resolution recovery and projected through the mid-point of the collimator. Finally, even though our emphasis was Cardiac SPECT, the dimensions of the detector might be useful for other applications such as preclinical SPECT[27] and clinical Brain SPECT.

**CONCLUSIONS**

We modeled in Monte-Carlo gamma-event interactions and scintillation light spread in a CsI hemi-ellipsoid detector and developed a robust algorithm that localized scintillation events to average 0.71(+/-0.44) mm within a hemi-ellipsoid CsI detector. Moreover, due to the high system magnification afforded by the crystal, the hemi-ellipsoid unit was capable of achieving <1 mm detector resolution in object space, assuming perfect pinhole collimator resolution recovery. Due to the high magnification in the central region of the hemi-ellipsoid detector, the average back-projection error in the mid-plane of the region of interest (object-space) was sub-mm at **0.4mm (+/-0.22)** mm. Thus, a hemi-ellipsoid detector will enable ultra-high resolution for Cardiac SPECT using collimator resolution recovery within reconstruction. In the future, a similar high curvature detector may be considered for Brain SPECT and small animal imaging.

**Acknowledgement**

The work presented here is in part of first-author H.R.Soysal's MSc thesis[28] (advisor by co-author J. D), towards his Medical Physics MSc degree, Department of Physics and Astronomy, LSU, Baton Rouge, LA. We thank committee member Professor Jeffrey Blackmon of Department of Physics and Astronomy, LSU for his feedback and discussions. We thank Bryce Smith for proof-reading the manuscript. Finally, we thank Louisiana State University High Performance Computing Center for providing the CPU hours to make this work possible.

**Reference**

1. *Radiobiology for the Radiologist*, E. J.Hall and A. J. Gaccia, 7th Ed, 2011

2. A. J. Einstein, K. W. Moser, R. C. Thompson, M. D. Cerqueira, M. J. Henzlova, "Radiation Dose to Patients From Cardiac Diagnostic Imaging", *Circulation.* vol.116, pp.1290-1305, 2007

3. W. L. Duvall, K. A. Guma, J. Kamen, L. B.Croft,M. Parides, T. George, M. J. Henzlova "Reduction in Occupational and Patient Radiation Exposure from Myocardial Perfusion Imaging: Impact of Stress-Only Imaging and High-Efficiency SPECT Camera Technology", *J Nucl. Med.* vol. 54, pp. 1251-1257, 2013

4. J. Dey, "Improvement of Performance of Cardiac SPECT Camera using Curved Detectors With Pinholes", *IEEE Trans. Nuclear Science*, vol.59, no.2,pp.334-347, April 2012

5. J. Dey and S.J. Glick, "SPECT Camera Design", *Patent No.,* US 8,519,351 B2, Aug 27, 2013

6. N. Bhusal, J. Dey et al, "Performance Analysis of a High-Sensitivity Multi-Pinhole Cardiac SPECT System with Hemi-Ellipsoid Detectors". *Med. Phys.* 46 (1), pp. 116-126, January 2019

7. L-E Adam, J.S. Karp, M. E. Daube-Witherspoon and R. J. Smith, "Performance of a Whole-Body PET Scanner Using Curve-Plate NaI(TI) Detectors", *J. Nucl. Med.*, vol. 42, no. 12, pp. 18201-1830, 2001

8. CardiArc: *www.cardiarc.com*

9. S. Agostinelli et al., "Geant4 – A Simulation Toolkit," *Nucl. Instrum. Meth.,* vol. A506, pp. 250-303, 2003.

10. M. C Maas, D. R. Schaart, D. J. van der Laan, P. Bruyndonckx, C. Lemaıtre, F. J Beekman and C. W E van Eijk,"Monolithic scintillator PET detectors with intrinsic depth-of-interactions correction", *Phys. Med. Biol.*, vol. 54, 2009, pp. 1893-1908

11. D. R Schaart, H. T van Dam, S. Seifert, R. Vinke, P. Dendooven, H. L¨ohner and F. J Beekman, "A novel, SiPM-array-based, monolithic scintillator detector for PET", *Phys. Med. Biol,* vol. 54, 2009, pp. 3501-3512

12. M. Morrocchi , G. Ambrosi, M. Giuseppina Bisogni, F. Bosi, M. Boretto, P. Cerello, M. Ionica, B. Liu, F. Pennazio, M. Antonietta Piliero, G. Pirrone, V. Postolache, R. Wheadon and A. Del Guerra, "Depth of interaction determination in monolithic scintillator with double side SiPM readout", *EJNMMI Physics*, vo. 4, 2017, pp. 1-25

13. D. J. van der Laan, D. R. Schaart, M. C. Maas, F. J. Beekman, P. Bruyndonckx and C. W E van Eijk, "Optical simulation of monolithic scintillator detectors using GATE/GEANT4", *Phys. Med. Biol.*, vol. 55, 2010, pp. 1659-1675

14. M. Bettiol, E. Preziosi, M.N. Cinti, C. Borrazzo, A. Fabbri, B. Cassano, C. Polito, R. Pellegrini and R. Pani, "A Depth-of-Interaction encoding method for SPECT monolithic scintillation detectors", *18th International Workshop on Radiation Imaging Detectors*, *JInstr, 10, C12054, 2016*

15. R. Pani, A.J. Gonzalez, M. Bettiol, A. Fabbri, M.N. Cinti, E. Preziosi,C. Borrazzo, P. Conde,R. Pellegrini, E. Di Castro and S. Majewski, "Preliminary evaluation of a monolithic detector module for integrated PET/MRI scanner with high spatial resolution", *16th International Workshop on Radiation Imaging Detectors*, *JInstr, 11, C06006, 2015*

16. J. W. Scrimger and R. G. Baker, "Investigation of Light Distribution from Scintillations in a Gamma Camera Crystal", *Phys. Med. Biol*, vol. 12, 1967, pp. 101-103

17. H.H. Barrett, W.C.J. Hunter, B.W. Miller, S.K. Moore, Y. Chen, and L.R. Furenlid, "Maximum-Likelihood Methods for Processing Signals From Gamma ray Detectors", *IEEE Trans. Nucl. Sci.*, 56(3), 725-735, 2009.

18. B.W. Miller, L.R. Furenlid, S.K. Moore, H.B. Barber, V.V. Nagarkar, and H.H. Barrett, "System Integration of FastSPECT III, a Dedicated SPECT Rodent-Brain Imager Based on BazookaSPECT Detector Technology", *IEEE Nucl. Sci. Symp. Conf. Record,* 4004-4008, 2009.

19. H.H. Barrett, L.R. Furenlid, H.B. Barber, and B.W. Miller, "Gamma camera including a scintillator and an image intensifier", United States Patent 7928397, Patent Issued 4/19/2011.

20. G. F. Knoll, Radiation Detection and Measurement, New York: John Wiley & Sons, 2010.

21. N. Grassi, G. Casini, M. Frosini, G. Tobia, M. Marchi, "PIXE characterization of CsI(Tl) scintillators used for particle detection in nuclear reactions," *Nuclear Instruments and Methods in Physics Research Section B, Beam Interactions with Materials and Atoms,* vol. 266, no. 10, pp. 2383-2386, 2008.

22. M. N. Polyanskiy, "Refractive index database," 01 Mar 2018. [Online]. Available: https://refractiveindex.info/?shelf=main&book=CsI&page=Li.

23. H. H. Li, "Refractive index of alkali halides and its wavelength and temperature derivatives," *Journal of Physical and Chemical Reference Data,* vol. 5, no. 2, p. 329–528, 1976.

24. F.-X. Gentit, "Litrani," 28 Sept 2007. [Online]. Available: https://crystalclear.web.cern.ch/crystalclear/LitraniX/Litrani/litrani/index.html

25. Saint-Gobain, "CsI(Tl) Thallium activated Cesium Iodide," 2007. [Online]. Available: https://www.crystals.saint-gobain.com/products/csitl-cesium-iodide-thallium.

26. J. B. Mosset, S. Saladino, J. F. Loude and C. Morel, Characterisation of arrays of avalance photodiodes for small animal positron emission tomography", Nucl. Instrum. Methods, Phys. Res. A. vol. 504, pp. 325-30

27. M. Carminati *et al*., "Validation and Performance Assessment of a Preclinical SiPM-Based SPECT/MRI Insert," in *IEEE Transactions on Radiation and Plasma Medical Sciences*, vol. 3, no. 4, pp. 483-490, July 2019, doi: 10.1109/TRPMS.2019.2893377.

28. H. R. Soysal, "Scintillation Even localization in Novel Hemi-ellipsoid Detector for SPECT", *MSc Thesis*, May 2019, Lousiana State University (advisor, Dr. J. Dey)